\documentclass[%
 reprint,twocolumn,
superscriptaddress,
 amsmath,amssymb,
 aps,
]{revtex4-2}

\usepackage{graphics}      
\usepackage{graphicx,float}      
\usepackage{longtable}     
\usepackage{url}           
\usepackage{bm}            
\usepackage[usenames]{color}
\usepackage{braket}
\usepackage[dvipsnames]{xcolor}
\usepackage{cancel}
\usepackage{soul}
\usepackage{comment}
\usepackage{amsmath,amssymb}
\usepackage{epstopdf}
\usepackage{ulem}
\definecolor{lightbrown}{RGB}{210, 180, 140}
\usepackage{footnote}
\usepackage[colorlinks=true,allcolors=blue]{hyperref}

\begin{document}

\title{Topological transitions in swarmalators systems}

\author{Patrick Louodop}
\affiliation{Research Unit Condensed Matter, Electronics and Signal Processing, University of Dschang, P.O Box 67 Dschang, Cameroon.}
\affiliation{MoCLiS Research Group, Dschang, Cameroon.}
\author{Michael N. Jipdi}
\affiliation{Quantum materials and Computing Group (QMaCG), Cameroon, P.O. Box 70, Bambili-Bamenda, Northwest Region, Cameroon.}
\affiliation{Department of Physics, Higher Teachers’ Training College, University of Bamenda, P.O. Box 11, Bamenda, Cameroon.}
\author{Gael R. Simo}
\affiliation{MoCLiS Research Group, Dschang, Cameroon.}
\affiliation{Laboratory of Electrotechnics, Automatics and Energy, Higher Technical Teachers, Training College (ENSET) of Ebolowa, University of Ebolowa, Cameroon}
\author{Steve J. Kongni}
\affiliation{Centre for Audio, Acoustics and Vibration, Faculty of Engineering and IT, University of Technology Sydney, Ultimo NSW 2007, Australia}
\affiliation{MoCLiS Research Group, Dschang, Cameroon.}

\author{Carmel Lambu}
\affiliation{Research Unit Condensed Matter, Electronics and Signal Processing, University of Dschang, P.O Box 67 Dschang, Cameroon.}
\affiliation{MoCLiS Research Group, Dschang, Cameroon.}
\author{Thierry Njougouo}
\email{thierry.njougouo@imtlucca.it}
\affiliation{IMT School for Advanced Studies Lucca, Lucca, Italy}
\affiliation{MoCLiS Research Group, Dschang, Cameroon.}

\author{Pablo D. Mininni}
\affiliation{Universidad de Buenos Aires, Facultad de Ciencias Exactas y Naturales, Departamento de Física, Ciudad Universitaria, 1428 Buenos Aires, Argentina.}
\affiliation{CONICET-Universidad de Buenos Aires, Instituto de Física Interdisciplinaria y Aplicada (INFINA), Ciudad Universitaria, 1428 Buenos Aires, Argentina.}
\author{Kevin O'Keeffe}
\affiliation{Starling Research Institute, Seattle, USA} 

\author{Hilda A. Cerdeira}
\affiliation{S\~ao Paulo State University (UNESP), Instituto de F\'{i}sica Te\'{o}rica, Rua Dr. Bento Teobaldo Ferraz 271, Bloco II, Barra Funda, 01140-070 S\~ao Paulo, Brazil.}
\affiliation{Epistemic, Gomez $\&$ Gomez Ltda. ME, Rua Paulo Franco 520, Vila Leopoldina, 05305-031 S\~ao Paulo, Brazil.}


\date{\today}

\begin{abstract}
After its development, the swarmalators model attracted a great deal of attention since it was found to be very suitable to reproduce several behaviors in collective dynamics. However, few works explain the transitions that are observed while varying system parameters. In this letter, we demonstrate that the changes observed in swarmalator dynamics are governed by changes in the system's topology. To provide a deeper understanding of these changes, we present a topological framework for the swarmalator system and determine the topological charge $Q$ and the helicity $\gamma$ of the corresponding topology. Investigations on synchronization and transition to synchronization are studied using this topological charge and the variance of the helicity. These techniques appear to offer a useful way to interpret some previously unexplained results and may help clarify the mechanisms involved in the transition to synchronization.

\end{abstract}

\maketitle

Collective dynamics has been among the most prominent and discussed topics during the last decades \cite{epid,hydro,neuro1,neuro2,louodop2017coherent}. To construct models that can reproduce real-life motions with fidelity, various investigations were reported during these years and applied to different problems \cite{louodop2017coherent,civil,prey1,prey2}. Within them, one can recall neuronal systems \cite{neuro1,neuro2,simo1,simo2}, prey-predators systems \cite{prey1,prey2,prey3,prey4,prey5}, mobile systems \cite{mobile,meli1,meli2,tanaka1,tanaka2} and more recently swarmalators \cite{swarm1,swarm2,swarm3}. In these studies, several behaviors have been found and linked to real phenomena such as coherent and incoherent dynamics \cite{swarm1}, chimera states \cite{chim1,chim2,chim3,chim4}, and the most investigated, which is synchronization \cite{syn1,syn2,syn3,syn4,syn5,syn6}. Synchronization is a fundamental phenomenon with significant implications across all fields of science \cite{syn1,syn2,syn3,syn4,syn5,syn6}. In the case of networks, it is one of the core phenomena leading to the emergence of collective behavior \cite{gomez}.

Introducing swarmalators in 2017, O'Keeffe et al.~\cite{swarm1} proposed a powerful tool to study collective motions and found several dynamical states, within which the basics are: (1) Static async, where the nodes are static in space with incoherent phases, (2) the static phase wave, where the phases rotate, (3) the splintered phase wave, where in addition to phase rotation the particles are spatially regrouped in clusters of almost identical phases, and (4) static sync, where the nodes are static in space and synchronize in phase (for more details, see Refs.~\cite{swarm1,swarm2,swarm3,swarm4,swarm5,kongni1,kongni2}). Several studies have been conducted on swarmalators, considering various situations such as pairwise and high-order interactions \cite{swarmnat}, or delayed interactions \cite{swarm6,lambu2026delay}, among others \cite{swarm7,swarm8,swarm9}.

Transitions to phase synchronization are typically examined using a Kuramoto-like order parameter (see Refs.~\cite{swarm1, swarm4, kongni1, kongni2}, where explosive and second-order phase transitions are reported). However, questions about the mechanism responsible for such transitions and their possible existence in other contexts remain unanswered. In this work, we demonstrate that the topology of the swarmalators is responsible for the dynamical changes leading to the transitions and synchronization. In particular, we propose a topological framework for swarmalator dynamics in the context of single- and double-layer systems (see Refs.~\cite{kongni1, kongni2}). We demonstrate that studying the system topology provides a comprehensive description of the transitions and routes to synchronization, allowing for careful analysis and characterization of the different regimes. The topological invariant $Q$ (topological charge, or winding number), as well as the helicity variance, play the role of synchronization witnesses; in particular, swarmalator textures are governed only by their winding number $Q$ and helicity $\gamma$.

Let us consider a two-dimensional swarmalator model, in which each agent is described by a two spatial and a single internal variable \cite{swarm1,kongni2}:
{\small
\begin{equation}\label{ee1}
{{\dot X}_i} = \frac{1}{N}\sum\limits_{j \ne i}^N {\left[ {\frac{{{X_j} - {X_i}}}{{\left. {\left| {{X_j} - {X_i}} \right.} \right|}}\left( {1 + J\cos \left( {{\theta _j} - {\theta _i}} \right)} \right)- \frac{{{X_j} - {X_i}}}{{{{\left. {\left| {{X_j} - {X_i}} \right.} \right|}^2}}}} \right]},
\end{equation}}
\begin{equation}\label{ee2}
  {{\dot \theta }_i} = w_i+\frac{K}{N}\sum\limits_{j \ne i}^N {\frac{{\sin \left( {{\theta _j} - {\theta _i}} \right)}}{{\left. {\left| {{X_j} - {X_i}} \right.} \right|}}},
\end{equation}
where $K$ stands for the phase coupling, $w_i$ are the natural phase frequencies of each element, and $J$ is the attraction or repulsion between the particles.  

The topological framework of the swarmalator system is based on the topological characteristics of the system and their implications for the emergence of global synchronization. To this end, we conduct an analysis of all swarmalators simultaneously, based on their respective internal variables, $\theta_i$. Thus, we associate each swarmalator with a Bloch sphere,

The Bloch sphere mapping serves as an auxiliary mathematical construction for topological classification, rather than a literal physical representation. Specifically, it is not intended to imply that swarmalators possess an additional spatial dimension. Instead, this mapping leverages the well-established theory of topological defects in vector fields on spherical manifolds (Bloch states), enabling us to evaluate the topological invariant of the collective states in our 2D system.
 Here, 2D swarmalator is characterised by its spatial position $(x_i, y_i)$ and internal phase $\theta_i$. In contrast, a  Bloch state on a sphere is described by position and two angular variables $(\theta_i, \phi_i)$, making a 3D-state space. The swarmalator's state can therefore be assimilated as a specific projection of a Bloch state onto the equatorial plane of the sphere.
The mapping $\Phi(\mathbf{r})$ transforms the spatial coordinates of each agent into a polar coordinate on a unit sphere, defined by the pair $(\phi(\mathbf{r}), \theta)$, where $\phi(\mathbf{r})$ is a prescribed radial function and $\theta$ is the agent's intrinsic phase. This constructs a synthetic Bloch vector for each agent on a virtual sphere \cite{sans2026}. 

This mapping helps us to define topological characteristics in terms of the swarmalators' internal phases.  Consequently, the full Bloch vector
    $\mathbf{s}_i~=~\big( \sin\phi(r_i)\cos\theta_i,\ \sin\phi(r_i)\sin\theta_i,\ \cos\phi(r_i) \big)$
  taken on the virtual sphere  reduces, under projection, to the effective 2D orientation vector
    $\mathbf{n}_i = (\cos\theta_i,\ \sin\theta_i,\ 0)$
defining swarmalators states via the internal variable $\theta_i$.
Under this description, the unit vector that describes the orientation of swarmalators' motion is $\left(n_x = \cos(\theta_i),n_y = sin(\theta_i)\right)$. 
This restricts our study to 2D topology, where we determine the topological invariants (the topological charges) which correspond to the winding number and the helicity. The winding number quantifies how the vector states of the elements rotate around a given singularity (vortex-like), and it is defined as follows (see supplemental SM I \cite{Supm}):
\begin{equation}\label{charge}
  Q = \displaystyle\frac{1}{2\pi}\oint_C \displaystyle n_x dy - n_y dx.
\end{equation}
The sign of Q  provides crucial information about the type of vortex identified by $Q=1$ for vortex-like, $Q= -1$ for anti-vortex-like, while $Q=0$ indicates a coherent state. 

In a swarmalator system (discretely distributed agents), the contour integral (Eq 3) is evaluated by defining an instantaneous elliptical contour around the particles' natural non-uniform spatial arrangement. The topological charge, Q, can therefore be calculated via the discrete analog of the contour integral, employing a Delaunay triangulation of the particle positions. ( see supplemental SM II \cite{Supm}).

On the other hand, the helicity $\gamma$ quantifies the vortex chirality. It can also be interpreted as the angle of the vortex rotation relative to a fixed axis. For a defined vortex centered in $\left(x_{0c}, y_{0c}\right)$, it can be defined as the reference angle for vortex rotation relatively to the radial axis $\overrightarrow{e}r = \cos(\phi)\overrightarrow{e}_x + \sin(\phi)\overrightarrow{e}_y$ where  $\phi = \arctan[(y - y_{0c})(x - x_{0c})]$. Thus, for a defined topology, the helicity is defined as:
\begin{equation}\label{charge}
  \gamma = \displaystyle\frac{\overrightarrow{n}\times \overrightarrow{e}_r}{\overrightarrow{n}\cdot \overrightarrow{e}_r}.
\end{equation}

For a system of N swarmalators, the local helicity $\gamma_i$ is calculated for
each particle $i$. The global helicity variance is then defined as the statistical variance of this set.
\begin{equation}\label{hvar}
   V(\gamma) =  \displaystyle\frac{1}{N} \sum\limits_{i = 1}^N \left( \gamma_i - \overline{\gamma}\right)^2,
\end{equation}
where $\bar{\gamma}$ is the mean helicity. The helicity variance $V(\gamma)$ is another crucial parameter that serves as a synchronisation indicator. It quantifies the disparity between the helicity of individual particles and the global helicity of the system, providing a direct measure of the spatial heterogeneity in the local alignment of Bloch-vector-states.

Complete synchronisation is obtained when both the winding number and the helicity variance vanish. Therefore, both charge and helicity variance behaviors are studied here to understand the system dynamics. 
As we will see, the topological charge and the helicity variance allow us to identify phase transitions in this system that because of their topological nature are not captured by a Kuramoto-like order parameter $R = \frac{1}{N}\displaystyle|\sum\limits_{i = 1}^N {{e^{j{\theta _i}}}}\displaystyle|$, with $j=\sqrt(-1)$.
Finally, in the study of swarmalators, the Phase Hamiltonian function also serves as a fundamental tool. Here, it can be approximated by \cite{kongni1,kongni2}.
\begin{equation}\label{hamil}
 H_{\text{phase}}=  - \sum \limits_{i \ne j}^N {\frac{K_{ij}}{{2N}}\cos \left( {{\theta _i} - {\theta _j}} \right)}.
\end{equation}

It provides an essential quantitative basis for comparing results across different models and parameter regimes. Furthermore, it serves as a robust order parameter for both phase synchronization and phase transitions, constituting the primary organizational principle underlying the observed state transitions. Although spatial dynamics such as particle motion in boiling phases are complex, the system's macroscopic state is determined by the collective phase order. The quantity $H_{\text{phase}}$ effectively quantifies this order, rendering it a relevant and reliable energy proxy, even in highly dynamic regimes where a full mechanical Hamiltonian would be intractable. Its minimization corresponds directly to the establishment of phase coherence, thereby driving transitions between distinct collective phases.

\begin{figure}
\centering
\begin{tabular}{cc}
\includegraphics[width=4cm,height=2.5cm,keepaspectratio]{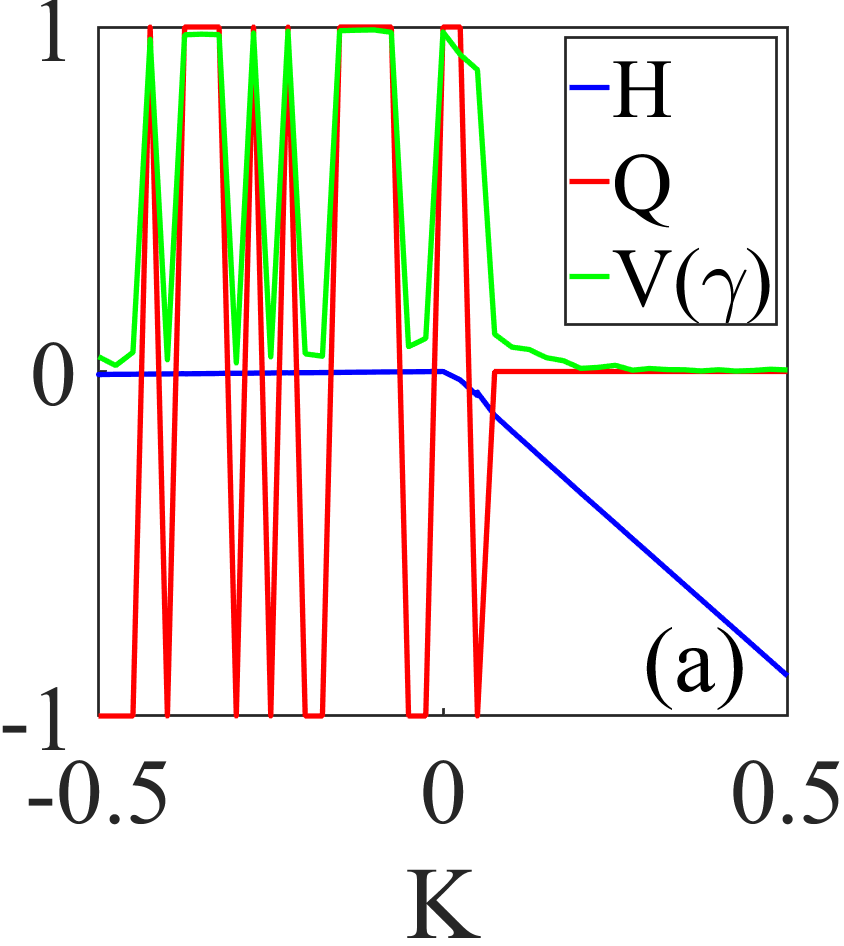}&
\includegraphics[width=4cm,height=2.5cm,keepaspectratio]{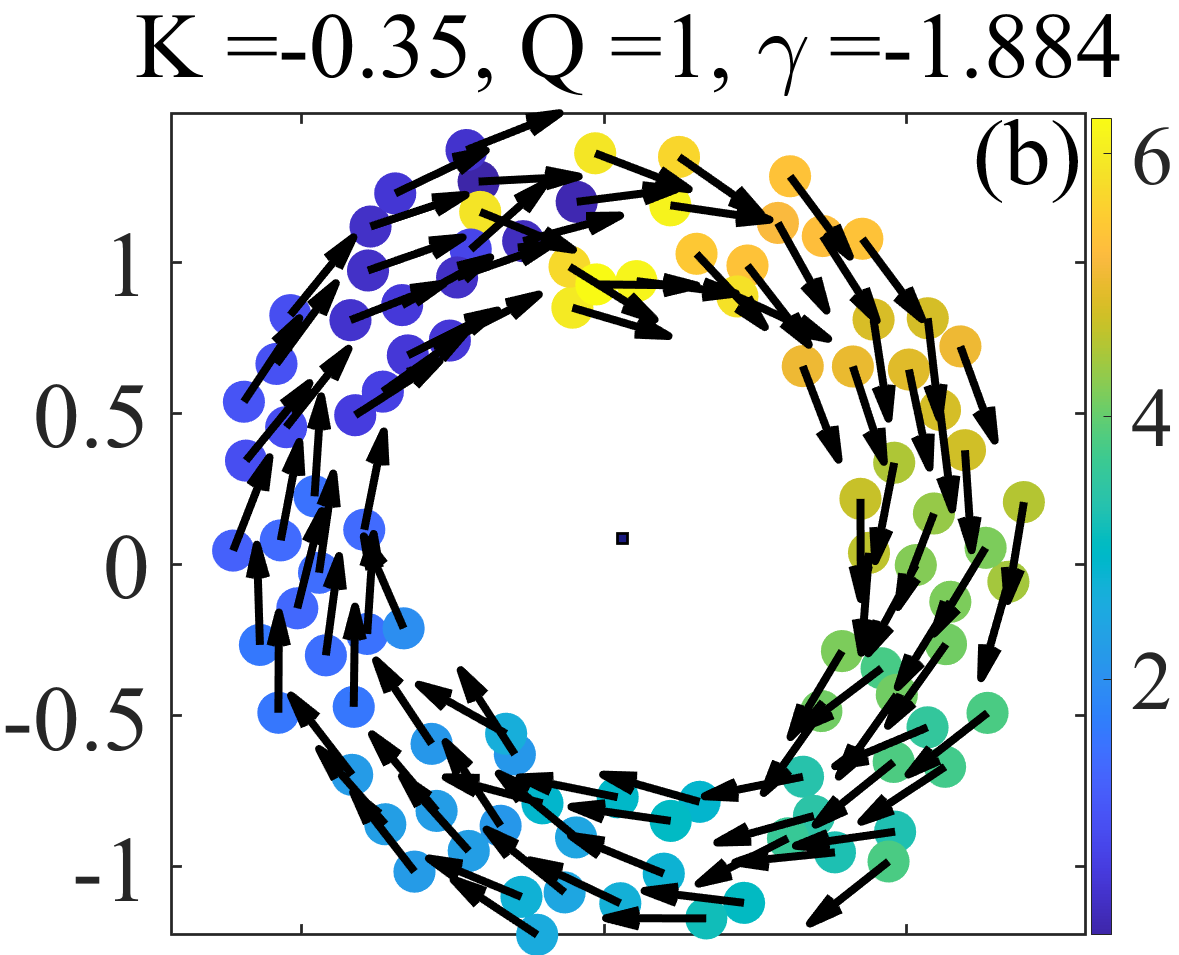}\\
\includegraphics[width=4cm,height=2.5cm,keepaspectratio]{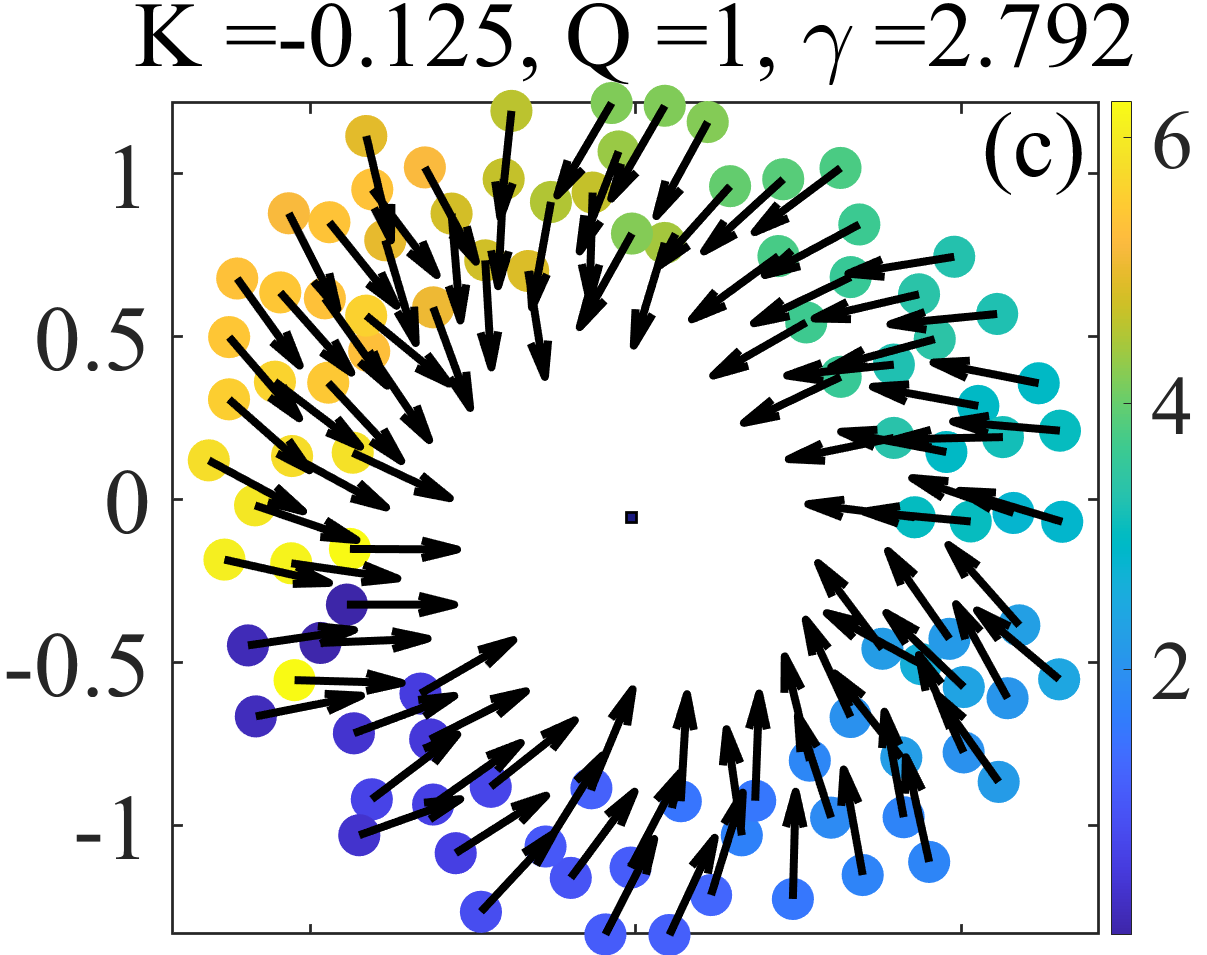}&
\includegraphics[width=4cm,height=2.5cm,keepaspectratio]{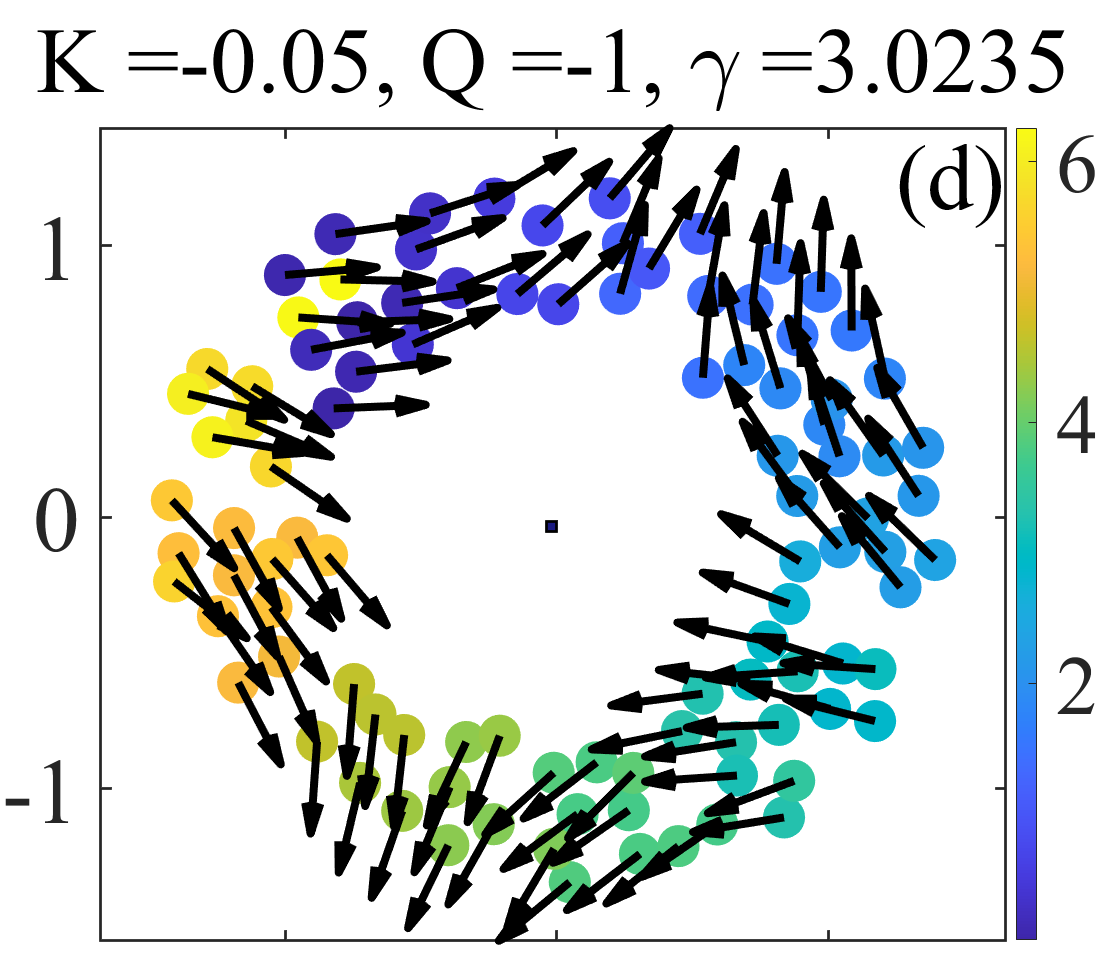}\\
\includegraphics[width=4cm,height=2.5cm,keepaspectratio]{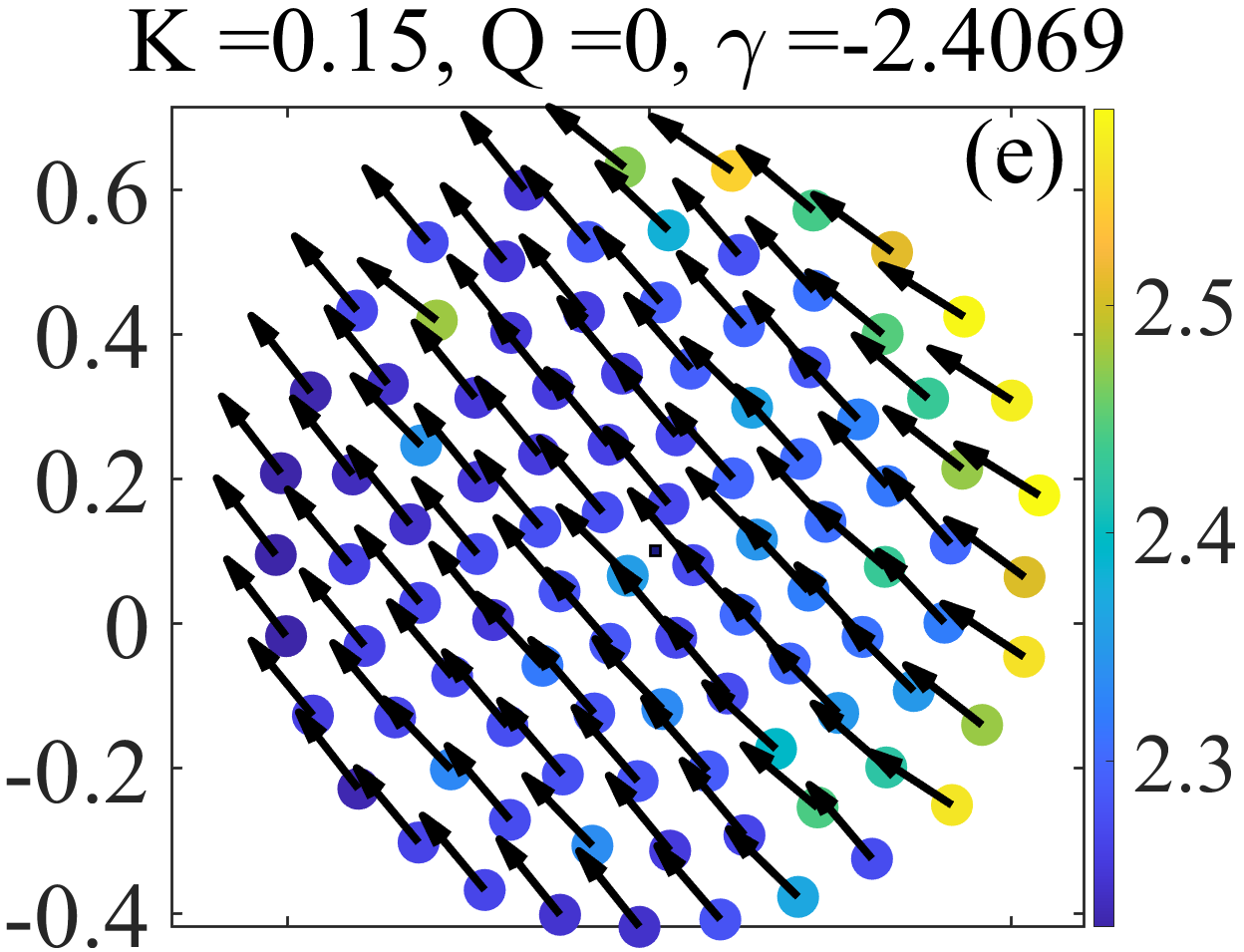}&
\includegraphics[width=4cm,height=2.5cm,keepaspectratio]{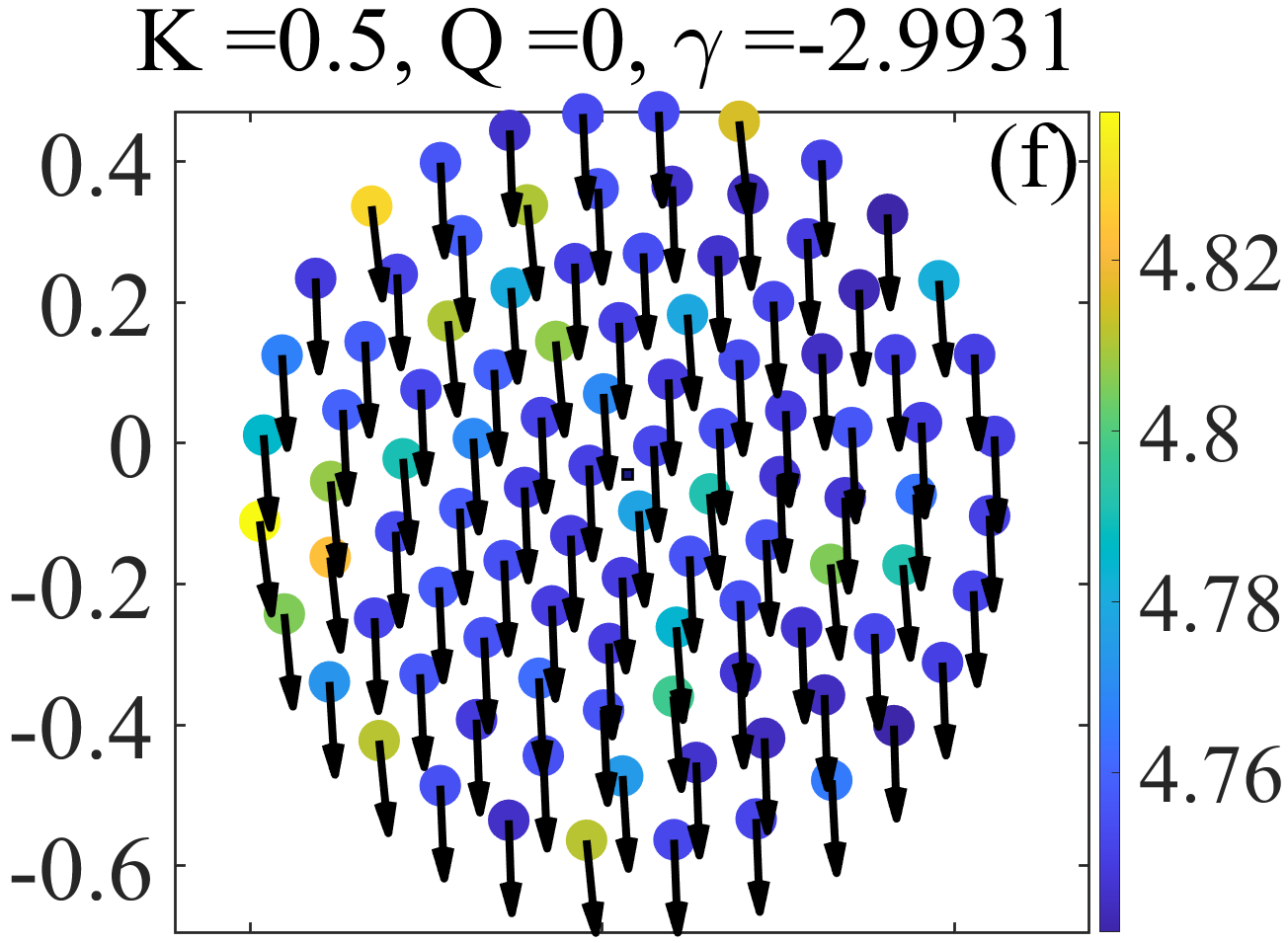}\\
\end{tabular}
\caption{Topological phase dynamics in the case of a second-order transition corresponding to the regime of Fig.~10 in Ref.~\cite{kongni2}. After reaching the critical coupling value, the topological charge is $Q=0$, meaning that the particles start to be or already are in a coherent state. The progressive tendency towards zero of the helicity variance,$V(\gamma)$, is a hallmark of the second-order transition, and when it reaches a value of zero, all elements of the network are perfectly synchronized (See Fig.~\ref{1lso}(a)). This analysis is proved by the color bars representing the values of the internal phases whose boundary values are shrinking while the systems are getting to synchrony (See Figs.~\ref{1lso}(b) - (f)). Also, as the arrows represent the projected “spin” direction of each swarmalator derived from the Bloch sphere mapping synchronization is characterized by their perfect parallel orientation.}
\label{1lso}
\end{figure}

We first present an analysis of a system comprising $N = 100$ swarmalators in a single layer. The system exhibits distinct phase-transition behavior depending on the distribution of the natural frequencies $\omega_i$ \cite{kongni2}. A second-order transition emerges under a mixed frequency distribution: $\omega_i = 0.005$ for $i \leq 30$, $\omega_i = 0.015$ for $31 \leq i \leq 61$, and randomly generated $\omega_i$ for $i \geq 62$ (see Fig.~\ref{1lso}).

The dynamics of the key order parameters, the topological charge $Q$ (red curve), the Hamiltonian energy $H$ (blue curve), and the helicity variance $V(\gamma)$ (green curve) are shown in Fig.~\ref{1lso}(a) as functions of the coupling strength $K$. The system undergoes a behavioral shift at a critical coupling $K_c$, where the energy $H = -K/2$ \cite{kongni2}. The sub-critical regime ($K < K_c$), where the topological charge has $|Q| = 1$, indicates a persistent vortex-like structure (recall that winding number is 1 for vortices and $-1$ for anti-vortices). This is visually confirmed by the heterogeneous helicity patterns in Figs.~\ref{1lso}(b)--(d). The helicity variance varies irregularly but remains greater than zero, signifying a failure in the local alignment of swarmalator orientations, and the system's energy $H$ remains constant at zero and is independent of $K$, denoting a stable, non-synchronized configuration.  In fact, the constant value of $H$ in Fig.~\ref{1lso}(a) for $K < K_c$ indicates that the system is in a stable or metastable state. Potential candidates for such states include topological phases, chimera states, synchronized states, or fragmented states. The behavior of the helicity variance with values significantly larger than zero confirms that the state is non-synchronized. Furthermore, the topological charge 
provides clear evidence of a topological phase. By combining these three analyses, we can conclude that the system is in a stable, non-synchronized configuration. 

In the super-critical regime ($K > K_c$), the system's energy progressively drops and remains at its theoretical minimum, $H = -K/2$, a value uniquely associated with a fully synchronized state. This progressive drop of the energy, which corresponds also to a progressive drop towards zero of the helicity variance, helps to identify the second order route to phase synchronization (this can be seen in Fig~.\ref{1lso}(e), where $Q=0$ while $V(\gamma) \neq 0$). The synchronized phase is finally reached through the collapse of both the topological charge and the helicity variance to zero. It represents a state of perfect synchronization with a trivial topology ($Q=0$ and $V(\gamma) = 0)$, as visualized in Fig.~\ref{1lso}(f). Thus, as $K$ is varied, the system undergoes a topological phase transition, marked by the discontinuous switch of the topological charge $|Q|$ from $1$ to $0$.

 Moreover, previous analysis of this model \cite{swarm1} has demonstrated that for certain values of the coupling parameter, swarmalators sharing almost the same internal phase are mutually attracted. This results in a spatial clustering of nodes with closely aligned phases, known as the splintered phase wave (SpPW) state. These results suggest that the SpPW state is due to the phase differences between distinct clusters.  Using our topological characterization of the system, we observe that each individual subgroup exhibits trivial topology. Consequently, swarmalators of similar phase aggregate into distinct groups, each behaving as a coherent collective unit characterized by a trivial topological charge and minimal variance in helicity. When considered collectively, the vector states of these groups give rise to well-defined topological configurations (vortex or anti-vortex structures) depending on the effective resultant vector state of each subgroup. It is therefore noteworthy that any cluster within a splintered phase wave state possesses the potential to form such topological structures. The details on the topological characterization of SpPW are found in supplemental SM III \cite{Supm}.

Next, we analyze a system comprising $N = 400$ swarmalators in a two-layer network (200 elements per layer) with $\omega_i = 0$ as in Ref.~\cite{kongni1} (Shown in supplemental SM IV \cite{Supm}). We focus our studies in the regime shown in Fig.~15 of this reference. It is observed that the system also exhibits a topological phase transition, characterized by the topological charge $Q$ evolving from large values, undergoing oscillations, and eventually saturating at $Q=0$, with $\max\{|Q|\} = 4$. We investigate this process by varying the inter-layer attractive coupling strength, which we denote by $\sigma_a$, with a step size $\Delta \sigma_a = 0.005$. The other parameters are fixed at a vision range of $r_c = 0.75$, intra-layer couplings $J_m = 0.1$ and $K_m = -0.1$ (for layer indices $m=1,2$). Random initial conditions for positions $X_i(0)$ and phases $\theta_i(0)$ are uniformly generated from the intervals $[-1, 1]$ and $[-\pi, \pi]$, respectively, where $i$ denotes the node index.

\begin{figure*}[htp!]
\centering
\begin{tabular}{c}
\includegraphics[width=15cm,height=7cm,keepaspectratio]{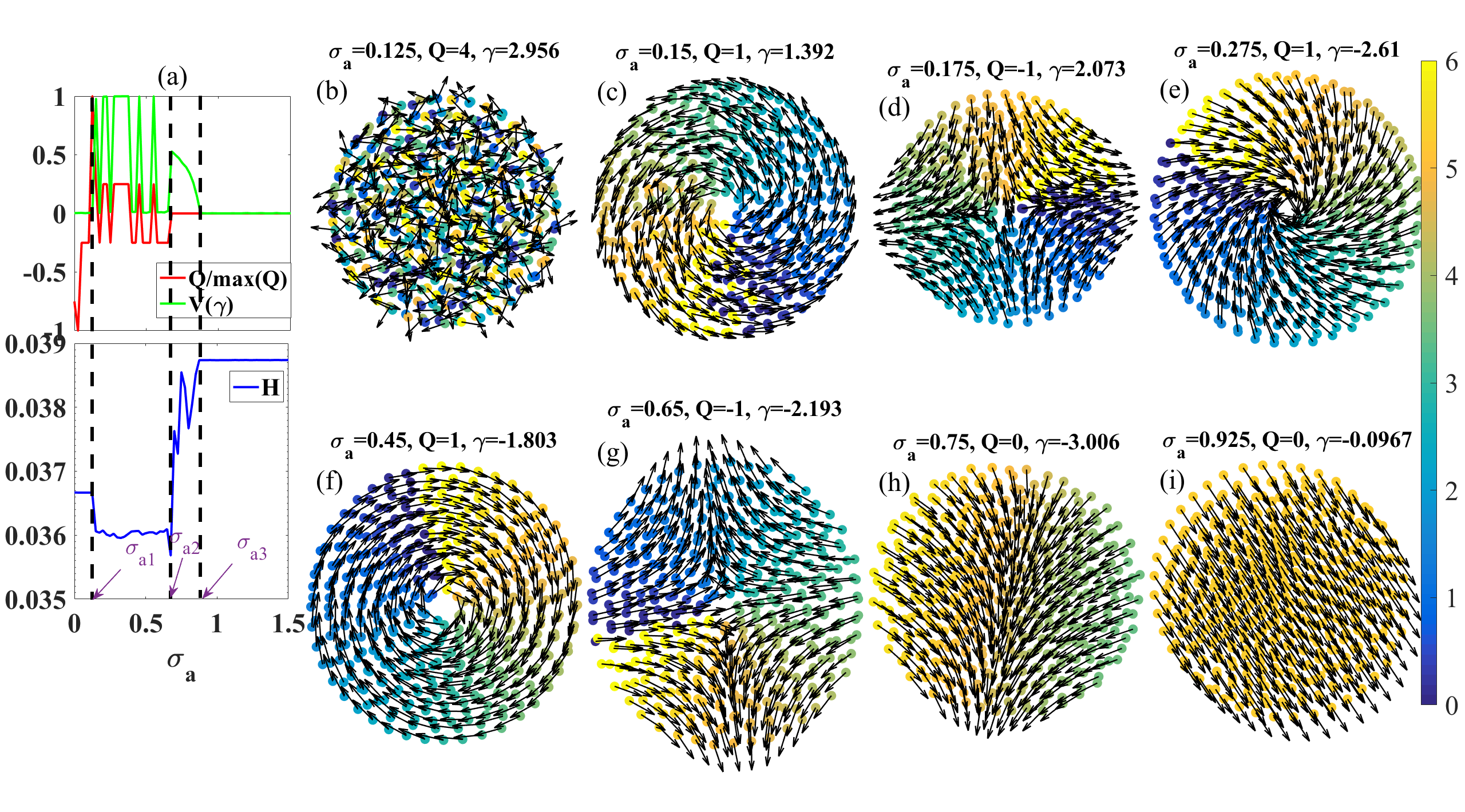}
\end{tabular}
\vspace{-0.5cm}
\caption{Topological phase dynamics corresponding to the regime studied in Fig.~15 of Ref.~\cite{kongni1}. The two energy transitions, i.e., the drop at $\sigma_{a1} \simeq 0.125$ (Figs.~\ref{2lso}(a) and (b)), and the jump at $\sigma_{a2} \simeq 0.65$(Figs.~\ref{2lso}(a) and (g)), find explanations in topological changes measured by the values of the winding number, as at the first transition it varies from $Q = 4$ (disorder state) at $\sigma_{a1}=0.125$ (Fig.~\ref{2lso}(b)) to $Q = 1$ (vortex state) at $\sigma_{a} = 0.15$ (Fig.~\ref{2lso}(c)), and in the second transition the charge varies from $Q = -1$ (anti-vortex) at $\sigma_{a2}=0.65$ (Fig.~\ref{2lso}(g)) to $Q = 0$ (trivial topology) at $\sigma_{a} = 0.675$ and later to synchronization (see Figs.~\ref{2lso}(h) and (i)). Vortex and anti-vortex can be seen in Figs.~\ref{2lso}(d) - (f).}
\label{2lso}
 \end{figure*}

This system undergoes a second-order phase transition, which can be divided into four distinct regimes (see Fig.~\ref{2lso}(a)), which we discuss in order of increasing coupling. {\it First regime:} A disordered multi-vortex regime ($\sigma_a < \sigma_{a1}=0.125$) characterized by low-coupling, where the inter-layer interaction is too weak to induce global order. The system's topology converges to a state resembling multiple vortices, analogous to a disordered, multi-clustered phase. Observations at the swarmalators boundary reveal groups of nodes rotating clockwise and counter-clockwise, as well as moving radially inward and outward. This phase juxtaposition results in a disordered texture and a topological charge with magnitude $|Q| > 1$. The pseudo-constant energy value in this phase suggests a metastable state (see Fig.~\ref{2lso}(b)),  which is justified by the fact that while the absolute value of the charge is $|Q|>1$ as in Fig.~\ref{2lso}(b), the energy remains quasi-constant.
{\it Second regime:} The coherent vortex-antivortex regime ($\sigma_{a1} < \sigma_{a} < \sigma_{a2}=0.65$), characterized by the intermediate coupling range where the system achieves a balance in which the energy drops and stabilizes at a new value, signaling a distinct phase. The swarmalators organize into a well-defined topological texture with $|Q| = 1$. The dynamics are characterized by irregular alternations between vortex-like (winding number $+1$) and anti-vortex-like (winding number $-1$) structures. The helicity dictates the swarmalators rotational direction, with a value of $-\pi/2$ conditioning a counter-clockwise rotation and $\pi/2$ a clockwise rotation (see Figs.~\ref{2lso}(d)--(g)). The helicity variance, while fluctuating, remains significantly non-zero, indicating a lack of global orientational order.
{\it Third regime:} An intermediate quasi-synchronized regime ($\sigma_{a2} < \sigma_{a} < \sigma_{a3}$) where elements are coherent but not synchronized, that is $Q = 0$ and $V(\gamma) \neq 0$, is obtained and defines the second order route to phase synchronization. This regime is shown on Fig.~\ref{2lso}(h) where elements are only locally quasi-synchronized.
{\it Fourth regime:} The synchronized trivial-topology regime ($\sigma_{a} > \sigma_{a3}$) is characterized by the strong coupling limit, where the interaction is sufficient to fully reorganize the system. Both the topological charge and the helicity variance collapse to zero. This signifies the achievement of perfect synchronization with a trivial topology, as shown in Fig.~\ref{2lso}(i). Again, as in the case with a single layer, the phase transitions are associated to topological changes.

\begin{figure}[htp!]
\centering
\begin{tabular}{c}
\hspace{-0.5cm}
\includegraphics[width=9cm,height=6cm,keepaspectratio]{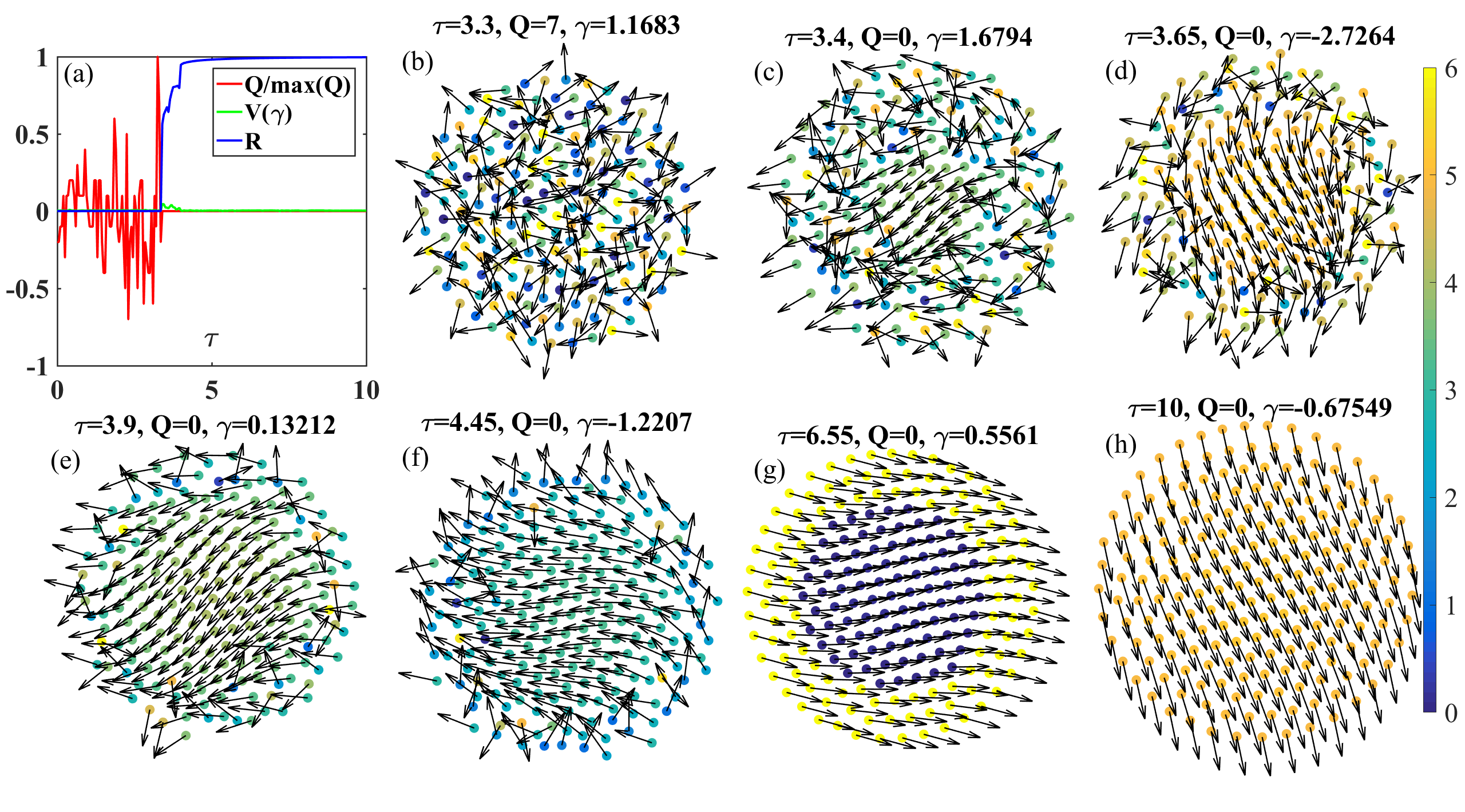}
\end{tabular}
\vspace{-0.5cm}
\caption{Topological phase dynamics in swarmalators with delayed interactions as considered in Ref.~\cite{swarm6}. Fig.~\ref{tauvar}(a) shows that the topological charge $Q$, the helicity variance $V(\gamma)$ and the order parameter $R$ are correlate to predict synchronization. Varying the delay $\tau$ leads the network from disorder (Fig.~\ref{tauvar}(b)) to a ring static sync state (See Figs.~\ref{tauvar}(f) - (h)) through the boiling state (See Figs.~\ref{tauvar}(c) - (e)), which is a non-topological chimera with $J = 0.1$ and $K = -0.5$. }
\label{tauvar}
\end{figure}

Finally, in a third study, we modify the system of Eqs.~(\ref{ee1}) and (\ref{ee2}) by introducing a delay in the phase dynamics as done in Ref.~\cite{swarm6,lambu2026delay}, with $N = 200$ swarmalators and $J = 0.1$. First, we investigate the behaviors of the swarmalators for fixed $K = -0.5$ as we vary the delay $\tau$. The system undergoes a phase transition when varying $\tau$, with the topological charge oscillating irregularly with $\max\{|Q|\} = 8$ and finally decaying to zero as $\tau$ increases, suggesting a topological phase transition ( see Fig.~\ref{tauvar}(a)). In fact, the swarmalators move from the async state (AS) to the ring static state (RSS) through the boiling state (BS) while varying the delay $\tau$  \cite{swarm6,lambu2026delay}. Considering the network texture, it seems that a central field organizes the nodes from the center to the border, whose effect increases with increasing $\tau$ (see Figs.~\ref{tauvar}(b) to (e)). In the figure, the RSS is visible in the concentric rings, where nodes in the same ring achieve synchronization but become phase-locked with those of the other rings even if the phase difference is very small (see Figs.~\ref{tauvar}(f) to (h)). This conclusion is further corroborated by the orientations of the swarmalators states, which visually represent the topological behavior of individual nodes in the Bloch sphere. The observed boiling state can thus be seen as a non-topological chimera state first because of the coexistence of coherent and incoherent regions and, secondly, because of the trivial topology they achieve.

\begin{figure}[htp!]
\centering
\begin{tabular}{c}
\hspace{-0.5cm}
\includegraphics[width=9cm,height=6cm,keepaspectratio]{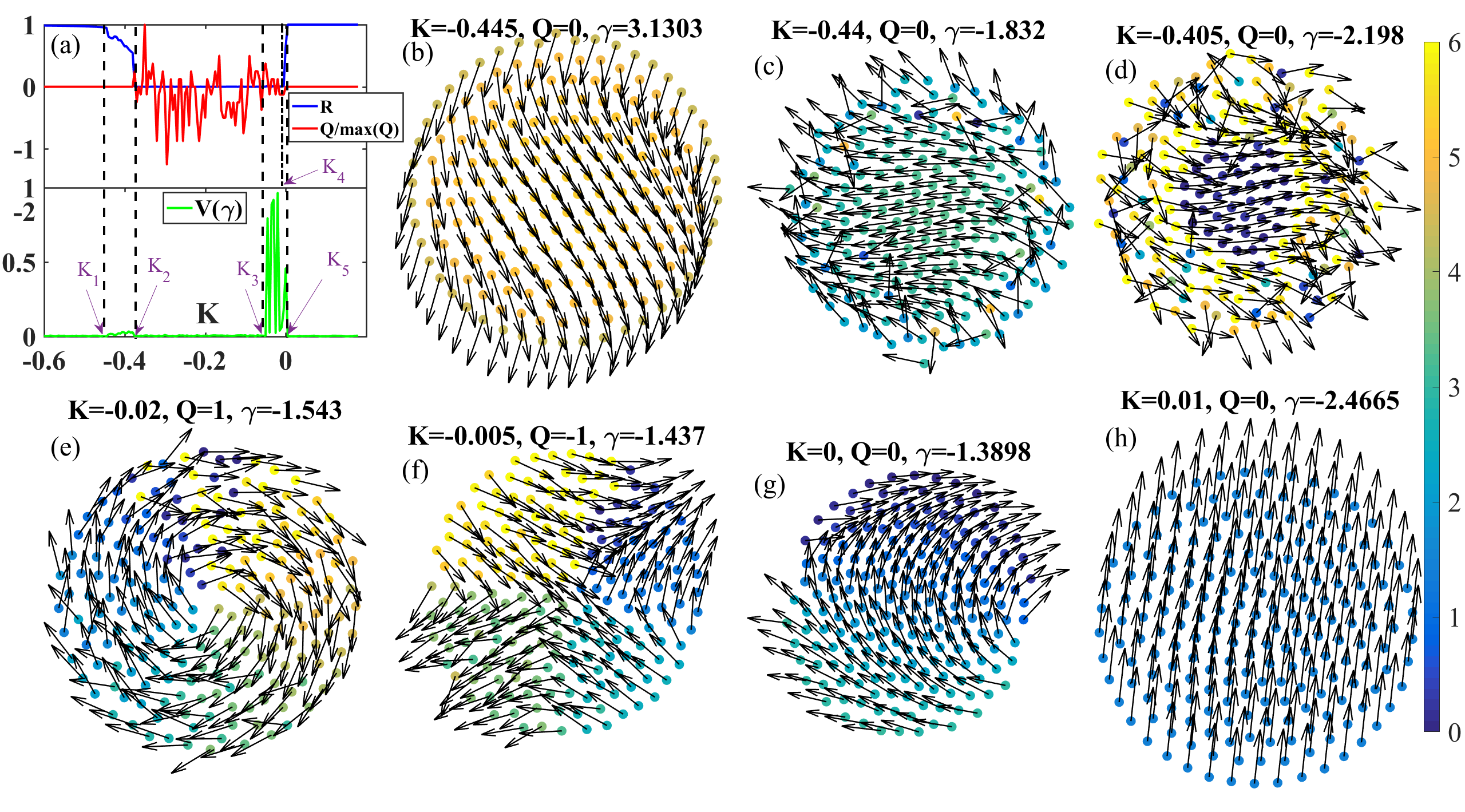}
\end{tabular}
\vspace{-0.5cm}
\caption{Topological transition in swarmalators with delayed interactions as considered in Ref.~\cite{lambu2026delay}, while varying the phase coupling $K$ for $\tau = 4.5$. Fig.\ref{kvar}(a) shows the six regions bounded by different values of the coupling parameter $K$ on the relative evolution of the order parameter $R$ (blue line), the topological charge $Q$ (red line) and the helicity variance $V(\gamma)$ (green line). The system moves from RSS (Fig.\ref{kvar}(b)) to coherence and later to a sync state (Figs.\ref{kvar}(g) and (h)) through disorder (Figs.\ref{kvar}(c) and (d)), followed by vortex-antivortex  topologies (Figs.\ref{kvar}(e) and (f)) that are non-detectable using the order parameter $R$ (blue curve in panel a).}
\label{kvar}
\end{figure}

Based on the values of the topological charge $Q$ (see Figs.~\ref{tauvar}(a) and (b)--(e)), there appears to be no formation of vortices or anti-vortices as the delay $\tau$ is varied for fixed $K$. However, this stands in stark contrast to the behavior observed when varying the coupling strength $K$ for a fixed value of $\tau = 4.5$. In this second case considered for swarmalators with delay, vortex formation follows the boiling state, while anti-vortices precede the system's organization into a coherent state (see results in Fig.~\ref{kvar}).

A comparison between Figs.~\ref{kvar}(b) to (h) clearly highlights the difference in topological texture. Figure~\ref{kvar}(b) (for $K = -0.445$) depicts the RSS characterized by a quasi-sync topology, while Fig.~\ref{kvar}(h) (for $K = 0.01$) shows the SS state. Both exhibit trivial topology. However, different topological states emerge between these two extreme states. Analysis of the topological charge $Q$ and helicity variance ${V}(\gamma)$ reveals that the system traverses six distinct phases (involving five phase transitions) as the coupling $K$ is varied. The classical order parameter, even with extensions, identifies only five phases, proving its inadequacy for characterizing these topologically distinct states. These are: (i) the RSS ($K < K_1$ ), (ii) a non-topological chimera state ($K_1 < K < K_2$), (iii) a disordered phase ($K_2 < K < K_3$), (iv) the vortex-antivortex phase ($K_3 < K < K_4$), (v) a quasi-sync state ($K_4 < K < K_5$), and (vi) the SS ($K > K_6$) (see Fig.~\ref{kvar}(a)).
Therefore, the complex changes induced by varying the coupling $K$ and delay $\tau$ are fundamentally topological in nature. The combined evolution of the topological charge and helicity variance provides the definitive signature for identifying the behavioral phases of the swarmalator system.

In conclusion, this work establishes that the phase transitions in swarmalator systems are fundamentally topological in origin.  The use of a synthetic Bloch sphere constitutes a strong auxiliary mathematical space where the constructed Bloch vectors are employed for topological analysis.  Employing the topological charge and helicity variance as robust topological witnesses, we demonstrated that varying the coupling strength induces distinct topological phase transitions in both single and multi-layer configurations. These topological metrics successfully quantify the system's energy landscape and global dynamics, revealing a state of pure synchronization at the parameter values $Q=0$ and $V(\gamma) = 0$. Furthermore, we identify the observed boiling states as a novel form of non-topological chimera, characterized by their trivial topology ($V(\gamma) \neq 0$). This topological framework proves essential as it overcomes the limitations of classical order parameters, which fail to distinguish complex phases such as the vortex-antivortex state.\\

$Acknowledgements$. HAC and PL thank the ICTP for hospitality. HAC acknowledges a FAPESP grant 2021/14335-0 for partial support. TN acknowledges support from the ``Reconstruction, Resilience and Recovery of Socio-Economic Networks'' RECON-NET - EP\_FAIR\_005 - PE0000013 ``FAIR'' -
PNRR M4C2 Investment 1.3, financed by the European Union
– NextGenerationEU.

\nocite{*}
\bibliographystyle{apsrev4-2}

\end{document}


\title{Supplemental Material:\\ Topological transitions in swarmalators systems}

\author{Patrick Louodop}
\affiliation{Research Unit Condensed Matter, Electronics and Signal Processing, University of Dschang, P.O Box 67 Dschang, Cameroon.}
\affiliation{MoCLiS Research Group, Dschang, Cameroon.}
\author{Michael N. Jipdi}
\affiliation{Quantum materials and Computing Group (QMaCG), Cameroon, P.O. Box 70, Bambili-Bamenda, Northwest Region, Cameroon.}
\affiliation{Department of Physics, Higher Teachers’ Training College, University of Bamenda, P.O. Box 11, Bamenda, Cameroon.}
\author{Gael R. Simo}
\affiliation{MoCLiS Research Group, Dschang, Cameroon.}
\affiliation{Laboratory of Electrotechnics, Automatics and Energy, Higher Technical Teachers, Training College (ENSET) of Ebolowa, University of Ebolowa, Cameroon}
\author{Steve J. Kongni}
\affiliation{Centre for Audio, Acoustics and Vibration, Faculty of Engineering and IT, University of Technology Sydney, Ultimo NSW 2007, Australia}
\affiliation{MoCLiS Research Group, Dschang, Cameroon.}
\author{Carmel Lambu}
\affiliation{Research Unit Condensed Matter, Electronics and Signal Processing, University of Dschang, P.O Box 67 Dschang, Cameroon.}
\affiliation{MoCLiS Research Group, Dschang, Cameroon.}
\author{Thierry Njougouo}
\email{thierry.njougouo@imtlucca.it}
\affiliation{IMT School for Advanced Studies Lucca, Lucca, Italy}
\affiliation{MoCLiS Research Group, Dschang, Cameroon.}

\author{Pablo D. Mininni}
\affiliation{Universidad de Buenos Aires, Facultad de Ciencias Exactas y Naturales, Departamento de Física, Ciudad Universitaria, 1428 Buenos Aires, Argentina.}
\affiliation{CONICET-Universidad de Buenos Aires, Instituto de Física Interdisciplinaria y Aplicada (INFINA), Ciudad Universitaria, 1428 Buenos Aires, Argentina.}
\author{Kevin O'Keeffe}
\affiliation{Starling Research Institute, Seattle, USA} 

\author{Hilda A. Cerdeira}
\affiliation{S\~ao Paulo State University (UNESP), Instituto de F\'{i}sica Te\'{o}rica, Rua Dr. Bento Teobaldo Ferraz 271, Bloco II, Barra Funda, 01140-070 S\~ao Paulo, Brazil.}
\affiliation{Epistemic, Gomez $\&$ Gomez Ltda. ME, Rua Paulo Franco 520, Vila Leopoldina, 05305-031 S\~ao Paulo, Brazil.}

\date{\today}

\begin{abstract}

\end{abstract}

\maketitle

\nocite{*}

\section*{SM I: Topological charge of 2D system}\label{app1}

Let us consider a vector field describing a topological texture. The topological charge of a defect is evaluated by \cite{Nagaosa2013,Jipdi2024}
\begin{equation}
Q = \frac{1}{4\pi} \iint \vec{n} \cdot \left( \frac{\partial \vec{n}}{\partial x} \times \frac{\partial \vec{n}}{\partial y} \right) dxdy.
\end{equation}
Here,
\begin{equation}
\begin{array}{l}
\vec{n} = \cos(\theta(x,y))\sin(\Phi(x,y)) \vec{i}\\ +
\sin(\theta(x,y))\sin(\Phi(x,y)) \vec{j} + \cos(\Phi(x,y) \vec{k},
\end{array}
\end{equation}
where $\vec{n}$ is the unit vector corresponding to the vector field. This formalism uses a spherical-like coordinate system to describe the unit vector field. Here, $\theta(x,y)$ is the azimuthal angle sweeping the entire $(x,y)$ plane, while $\Phi(x,y)$ is the polar angle, which captures the inclination of the field vector from the z-axis and thus describes the variation of its z-component across the plane. 

Based on these components, the topological charge is expressed in terms of $\theta$ and $\Phi$ as
\begin{equation}
\begin{array}{l}
Q = \frac{1}{4\pi} \iint \sin(\Phi(x,y)) U\left( {x,y} \right) dxdy,\\ \textrm{with} \\
U\left( {x,y} \right) = \left( {\frac{{\partial \Phi (x,y)}}{{\partial x}}\frac{{\partial \theta(x,y)}}{{\partial y}} - \frac{{\partial \Phi (x,y)}}{{\partial y}}\frac{{\partial \theta(x,y)}}{{\partial x}}} \right).
\end{array}
\end{equation}

It is instructive to point out that the polar angle $\Phi(x,y)$ depends solely on the radial distance while the azimuthal angle $\theta(x,y)$ is a function of the geometric azimuthal angle in the plane. Based on that we can rewrite this expression as
\begin{equation}
Q = \frac{1}{4\pi} \iint \sin(\Phi(r)) \frac{d\Phi(r)}{dr} \frac{d\theta(\phi)}{d\phi} \left( \frac{\partial r}{\partial x} \frac{\partial \phi}{\partial y} - \frac{\partial r}{\partial y} \frac{\partial \phi}{\partial x} \right) dxdy .
\end{equation}

In the latter expression, the factor $\left( \frac{\partial r}{\partial x} \frac{\partial \phi}{\partial y} - \frac{\partial r}{\partial y} \frac{\partial \phi}{\partial x} \right)$ is the Jacobian for the transformation from polar to Cartesian coordinates. The topological charge in polar coordinates is therefore given by:
\begin{equation}
Q = \frac{1}{4\pi} \iint \sin(\Phi(r)) \frac{d\Phi(r)}{dr} \frac{d\theta(\phi)}{d\phi} dr d\phi.
\end{equation}

From this relation, we can see that the radial and azimuthal parts are separable. This helps us split the integral into two parts,
\begin{equation}
Q = \frac{1}{4\pi} \left[ \int \sin(\Phi(r)) d\Phi(r) \right] \int d\theta(\phi).
\end{equation}

The radial part can be easily evaluated since the behavior of the z-component is known,
\begin{equation}
\int \sin(\Phi(r))d\Phi(r) = -\cos\Phi(r)\bigg|_{0}^{\pi} = 2.
\end{equation}
Therefore,
\begin{equation}
Q = \frac{1}{2\pi}\int d\theta(\phi) \quad \text{or} \quad Q = \frac{1}{2\pi}\oint (\vec{\nabla}\theta) \vec dl,
\end{equation}
where $l$ here maps the elliptic contour encircling all swarmalators. \\
This provides an expression for the topological charge of a texture projected onto the $(x,y)$ plane. Therefore, for any 2D vector field described by a unit vector $\vec{n} = (n_x = \cos(\theta), n_y = \sin(\theta))$, we can associate a hypothetical polar angle $0 < \Phi(x,y) < \pi$ such that the texture matches the definition in formula (A 3). Consequently, the topological charge can be evaluated simply as:
\begin{equation}
Q = \frac{1}{2\pi}\oint (\bar{\nabla}\theta).\vec dl .
\end{equation}

This expression can also be written in terms of $n_x$ and $n_y$ for generalization. Given any 2D vector $\vec{n} = (n_x, n_y)$ the associated angle is $\theta = \arctan\left(\frac{n_y}{n_x}\right)$ thus
\begin{equation}
d\theta = \frac{d\left(\frac{n_y}{n_x}\right)}{1 + \left(\frac{n_y}{n_x}\right)^2} = \frac{n_xdn_y - n_ydn_x}{n_x^2 + n_y^2} .
\end{equation}
Thus the generalized formula for unit vectors
\begin{equation}
Q = \frac{1}{2\pi}\oint n_xdn_y - n_ydn_x .
\end{equation}
\\
\section*{SM II : Brief introduction to the calculation of the topological charge}\label{app2}

To compute the topological charge $Q$ defined by the contour integral
\begin{equation}
Q = \frac{1}{2\pi}\oint \bar{\nabla}\theta .\vec dl
\end{equation}
we implement the following numerical procedure:\\
1. \textbf{Contour construction}: For each instantaneous configuration, we compute the
Delaunay triangulation of the particle positions. From this triangulation, the convex
hull of the particle set is extracted, defining a piecewise-linear closed contour $C$ that
tightly encloses the swarmalator cluster.\\
2. \textbf{Discrete phase gradient}: Along each directed edge $\braket{i,j}$ of the hull contour, we compute the minimal phase difference
$$\Delta\theta_{ij} = \theta_{j} -  \theta_{i} \mod{2\pi}$$
ensuring that  $\Delta\theta_{ij} \in \left[-\pi, \pi\right]$\\
3.\textbf{Integral summation}: The line integral is approximated as the sum of these discrete contributions:
\begin{equation}
Q_{num} = \frac{1}{2\pi}\sum_{\substack{\braket{i,j} \in C} }\Delta\theta_{ij}.   
\end{equation}
This Delaunay-based approach robustly handles non-uniform particle distributions and provides a discrete analog of the continuous contour integral. The method is standard in particle-based computations and topological invariants (see \cite{scg1997,Perumal2019}).\\
 4. \textbf{Boundary conditions}: The simulations were performed using Fixed Boundary Conditions (FBC). For each instant, the integration contour $C$ is defined as an ellipse that encloses all particles. This ellipse is determined by fitting parameters (position, orientation, axis lengths) so that it best fits the shape of the instantaneous spatial distribution of the swarmalator cluster using a principal component analysis (PCA) of particle positions. The semi-major and semi-minor axes are set to $1.2$ times the corresponding standard deviations obtained from the PCA, ensuring the contour encloses all particles with at least one particle lying on or near the contour path for numerical stability.
 
\section*{SM III : Splintered Phase Wave (SpPW)} \label{app4}

The splintered phase wave state is characterized by a cluster formation of nodes with very close phase values. However, the difference in color between clusters means that the difference in phases of two groups is significantly large. Considering Eqs.1 and 2 of the main text with $w_i = 0.005$, $J=1$ and $K=-0.1$ for $N=600$ nodes, we evaluate the topological charge $Q$, the helicity $\gamma$ and the helicity variance $V(\gamma)$and obtain: $Q=1$, which means that the SpPW state is a vortex-like dynamics, and $V(\gamma)\neq 0$ implies the absence of synchrony.  

\begin{figure}[htp!]
    \centering
    \includegraphics[width=0.25\textwidth]{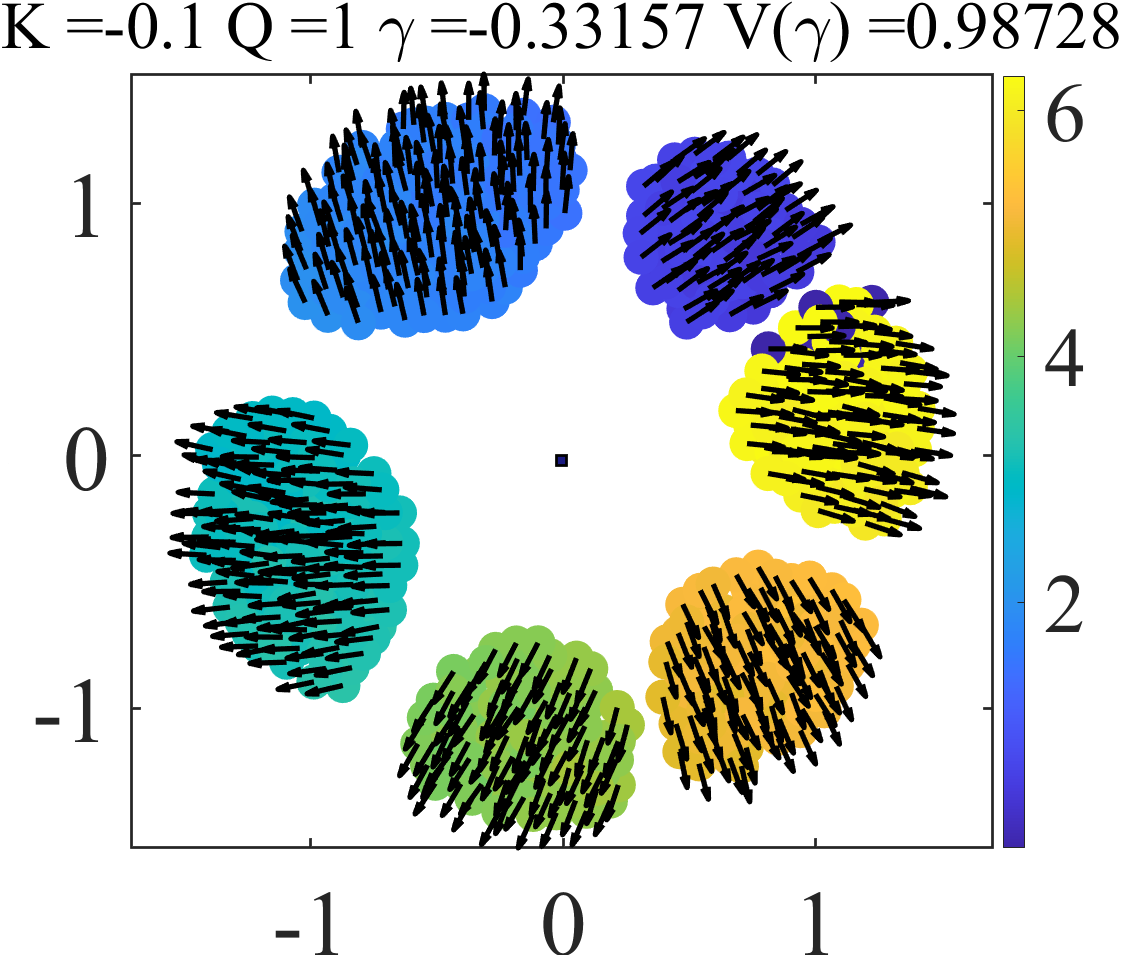}
    \caption{Splintered phase wave observed from Eqs.1 and 2 of the main text with $w_i = 0.005$, $J=1$ and $K=-0.1$ for $N=600$ nodes presenting a vortex - like behavior. }
    \label{SpPW}
\end{figure}

\section*{SM IV : Multilayered Model Used} \label{app3}

While the space motions, in both layers, are still described by Eq.~1 (see the main text), the phase equations in Eq.~2 are modified as follows (see \cite{Kongni2023}):
\begin{equation}
 \dot \theta _i^m = \underbrace {\frac{{{K_m}}}{N}\sum\limits_{j \ne i}^N {\frac{{\sin \left( {\theta _j^m - \theta _i^m} \right)}}{{\left| {X_j^m - X_i^m} \right|}}} }_{{\rm{intra\ layer\ dynamics}}} + C_{{\rm{att}}}^m(X_{ij}^m,\theta _{ij}^{ml}) + C_{rep}^m(X_{ij}^m,\theta _{ij}^{ml}),
\end{equation}
where $m = 1, 2$ identifies the layer, $\theta_i^m$ represents the internal phase of element $i$ in layer $m$, $\theta_{ij}^{ml} = \theta_i^m - \theta_j^l$, and $X_{ij}^m = X_j^m - X_i^m$,
\begin{equation}
C_{\text{att}}^m = \underbrace{\frac{\sigma_a}{N_a^m} \sum_{j \neq i}^N F_a^l (i, j) \frac{\sin \left(\theta_j^l - \theta_i^m\right)}{D_{ij}^m}}_{\text{Attractive inter layer dynamics}},
\end{equation}
\begin{equation}
C_{\text{rep}}^m = \underbrace{\frac{\sigma_r}{N_r^m} \sum_{j \neq i}^N F_r^l (i, j) \frac{\sin \left(\theta_j^l - \theta_i^m\right)}{D_{ij}^m}}_{\text{Repulsive inter layer dynamics}}
\end{equation}
where $\sigma_a > 0$ is the inter-layer coupling strength for attractive interaction. $N_a^m$ is the number of elements with attractive interaction inside the vision range in layer $m$, and $N_r^m$ is the number of those in the same layer that are not inside the vision range. $F_a^l$ is the $N \times N$ matrix representing the attractive inter-layer interaction. It shows the connectivity (attractive ) created by the $i^{th}$ element with those in the opposite layer:
\begin{equation}
\begin{aligned}
F_a^l (i, j) &= 
\begin{cases} 
1 & \text{if } j^{(l)} \in \Delta_{i(m)}^{(l)} [r_c^{(l)}], \\ 
0 & \text{otherwise} 
\end{cases}
\end{aligned}
\end{equation}
\begin{equation}
\begin{aligned}
F_r^l (i, j) &= 
\begin{cases} 
1 & \text{if } j^{(l)} \notin \Delta_{i(m)}^{(l)} [r_c^{(l)}], \\ 
0 & \text{otherwise} 
\end{cases}
\end{aligned}
\end{equation}
where $l = (1, 2)$, $m = (1, 2)$, and $\Delta_{i(m)}^{(l)} [r_c^{(l)}]$ is the set of elements inside the domain [vision range $r_c^{(l)}$] created around the projection of the $i^{th}$ element of layer $m$ onto layer $l$.

\bibliographystyle{apsrev4-2}
